\documentstyle[psfig]{espart}
\begin{document}
\begin{frontmatter}
  \title{{\small J. Stat. Phys., Vol. 86, 421-429 (1996)}\\[1cm]
A Statistical Approach to Vehicular Traffic} 
\author[HUB]{Jan Freund\thanksref{bylinejanf}}
\author[HUB,CHIC]{Thorsten P\"oschel\thanksref{bylinetp}}

  \address[HUB]{Humboldt--Universit\"at zu
    Berlin, Institut f\"ur Physik, Unter den Linden 6, D--10099 
    Berlin, Germany, 
    http://summa.physik.hu-berlin.de:80/$\sim$thorsten/}
  \address[CHIC]{The James Franck Institute, The University of Chicago, 
    5640 South Ellis Av., Chicago, Illinois 60637}
  \thanks[bylinetp]{E--mail: thorsten@trex.uchicago.de}
  \thanks[bylinejanf]{E--mail: freund@physik.hu-berlin.de}

---------
\date{\today}
\maketitle
\begin{abstract}
A two--dimensional cellular automaton is introduced to model the flow
and jamming of vehicular traffic in cities. Each site of the automaton
represents a crossing where a finite number of cars can wait
approaching the crossing from each of the four directions.  The flow
of cars obeys realistic traffic rules. We investigate the dependence
of the average velocity of cars on the global traffic density. At a
critical threshold for the density the average velocity reduces
drastically caused by jamming.  For the low density regime we provide
analytical results which agree with the numerical results.
\end{abstract}
\end{frontmatter}
\section{Introduction}
Vehicular traffic on highways as well as in cities tends to suffer
from a jamming transition when the global traffic density exceeds a
critical threshold value. The phenomena related to traffic jams have
attracted the attention of engineers and physicists since many years
and there exists a large variety of experimental observations,
see~\cite{Treiterer:1975,MushaHiguchi:1976,May:1993,Leutzbach:1988}
and references therein. The formation of traffic jams has been
investigated by many authors using various methods. As early as 1955
Lighthill and
Whitham~\cite{LighthillWhitham:1955a,LighthillWhitham:1955b} described
the spontaneous formation of regions of increased car concentrations
and shock waves using their approach of kinematic waves, i.e. wave
motion where the spatial value of the flow is a function of the
spatial concentration distribution. Prigogine and
Herman~\cite{PrigogineHerman:1971,HermanPrigogine:1979ug} investigated
traffic jams on highways by performing a stability analysis of
hydrodynamic equations. One--dimensional cellular automata
models~\cite{Wolfram:1986BOOK} for the simulation of traffic flow have
been proposed e.g.
in~\cite{NagelSchreckenberg:1992,Wiedemann:1974,Nagatani:1993b}.  This
type of models has been studied intensively using massively parallel
computers~\cite{NagelHerrmann:1993,Nagel:1994,Nagel:1994a,NagelSchleicher:1993,NagelRasmussen:1994}.
Schadschneider and
Schreckenberg~\cite{SchadschneiderSchreckenberg:1993} solved a 
one--dimensional cellular automaton model analytically in the mean field
approximation, Csah\'ok and Vicsek~\cite{CsahokVicsek:1994}
investigated their automaton in the presence of quenched noise.
Nonlinear wave descriptions have been proposed in
\cite{Walker:1982,Sick:1988,KernerKonhaeuser:1993}. Ben--Naim et. al
\cite{BenKrapivskyRedner:1994} apply a ballistic aggregation process
to model the kinetics of clustering in one--dimensional traffic
flow. Hydrodynamic approaches have been investigated by various
authors, e.g.~\cite{Helbing:1992,Helbing:1994}. These models are based
on certain assumptions concerning the ``hydrodynamic'' properties of
traffic flows such as the velocity versus density
relation~\cite{Kuehne:1984,Kuehne:1988}, and the viscous
terms~\cite{KernerKonhaeuser:1993}.

One--dimensional traffic has been studied extensively and the developed
models are sufficiently sophisticated to reproduce experimental
observations, i.e. the fundamental diagram (throughput versus flow
density) of one lane roads. Some of the authors
(e.g.~\cite{TakayasuTakayasu:1993}) claim that the size distribution
of the jams obeys a power law, but very large scale computations have
shown that the lifetimes of the jams and hence their sizes reveal a
characteristic cut--off which seems to be no finite size effect of the
simulation~\cite{Nagel:1994}. There are several models describing
traffic flow in more than one dimension. The analysis of traffic on a
network with multiple sources and sinks can be found e.g.~in
\cite{Sheffi:1985,MahmassaniJayakrishnanHerman:1990,HilligesReinerWeidlich:1993,Haberman:1977}.
Nagatani~\cite{Nagatani:1993} describes a cellular automaton
representing a two--lane roadway.

An interesting new approach was currently proposed by Bando et
al.~\cite{BandoEtAl:1995} who express the rules which govern the
behavior of the cars (i.e. acceleration and deceleration) by a dynamic
equation for each car $\ddot{x}_n = \alpha\left\{
V\left(x_{n+1}-x_{n}\right)-\dot{x}_n\right\}$. The interaction of
neighboring cars is expressed in terms of the function $V$ of the
distance between the cars. Using this approach the authors connect in
some respect the model of the cellular automata with molecular
dynamics.

Two--dimensional cellular automata designed to simulate traffic in a
city have been investigated by some authors. Biham et
al.~\cite{BihamMiddletonLevine:1992} proposed a three state cellular
automaton model where at a given time step each site can be occupied
by a car moving from South to North, by a car moving from West to
East, or the site may be empty. The time behavior is ruled by
synchronous traffic lights at each site, allowing alternatively for
vertical or horizontal traffic. Obviously there are situations when
the traffic jams, i.e. when the traffic light allows for driving but
the next site is occupied. The model given
in~\cite{BihamMiddletonLevine:1992} and particularly the spatial
correlations in the jamming phase have been numerically investigated
by Tadaki and Kikuchi~\cite{TadakiKikuchi:1994}. A similar model but
with ``faulty'' traffic lights was studied
in~\cite{ChungHuiGu:1995}. Nagatani~\cite{Nagatani:1993a} investigated
the spreading of a jam which is induced by an accident using an
extremely simple automaton rule. Numerically he finds scaling laws for
the size of the spreading jam as a function of time elapsed since the
occurrence of the accident which cannot be derived analytically so
far. Fukui et al.~\cite{FukuiIshibashi:1993} numerically investigated
the evolution of ensemble averages of the jamming process for the
simple automaton model described
in~\cite{BihamMiddletonLevine:1992}. For the same
model~\cite{BihamMiddletonLevine:1992} Chau et
al.~\cite{ChauHuiWoo:1995} recently gave an analytic upper bound
(depending on the dimension) for the critical car density $\eta_{cr}$,
i.e.~when 
the system transits from the ``moving phase'' into the ``jamming
phase''. In two dimensions they found $\eta_{cr}\le 11/12$. Obviously
this upper bound is not close to the value which was observed in
numerical simulations. In the case of Nagatani and
Seno~\cite{NagataniSeno:1994} there is a set of parallel one--way
streets, all oriented in the same direction ($x$--direction), which is
intersected by a single perpendicular one--way street in
$y$--direction.

For fixed car density on the perpendicular street $\rho_y$ they find
that the flux of cars $J_x$ rises linearly with the car density
$\rho_x$ until a characteristic critical threshold $\rho_x^c$ is
reached. When further increasing the density $\rho_x$ the flux $J_x$
approximately remains constant while the average velocity of the cars
$\langle v_x \rangle$ drops. This rather sharp change in the overall
behavior is due to the formation of traffic jams where the flow is
irregular and discontinuous. When further increasing the density
$\rho_x$ the discontinuous character of the flow disappears at a
second threshold $\rho_x^C$ and the flux $J_x$ declines linearly. This
behavior corresponds to the symmetry of the car density in
$y$--direction and the spaces between the cars $1-\rho_y$. Indeed, one
finds that the critical densities $\rho_x^c$ and $\rho_x^C$ are almost
exactly symmetric, e.g. for $\rho_y=0.3$ one finds $\rho_x^c=0.31$ and
$\rho_x^C=0.69$.

Cuesta et al.
\cite{CuestaMartinezMoleraSanchez:1993,MoleraMartinezCuesta:1994} have
been the first who reported on simulations of a two--dimensional
automaton where the cars are allowed to change their direction.  In
their model each crossing can be occupied by one car or it can be
empty. Each car is assigned a preferred direction; the parameter
$w_i(r)$ is the probability that the $i$--th car at the position $r$
moves in the next time step horizontally and $1-w_i(r)$ is the
probability to move in vertical direction. Horizontal motion is
allowed at even time steps, vertical motion at odd times. There have
been defined two variants, A: there are only one--way streets directed
from South to North and from East to West. Half of the cars is given
the trend $w_i(r)=\gamma$ and the other half is given
$w_i(r)=1-\gamma$. In the second variant B there are one--way streets
pointing alternatively South $\rightarrow$ North and North
$\rightarrow$ South, and West $\rightarrow$ East and East
$\rightarrow$ West, respectively. The trends of the cars which are
subdivided into four equal groups, point into one of the four possible
directions.

We should mention that several authors claim that at least the 
one--dimensional traffic flow problem is closely related to sand flowing in
pipes,
e.g.~\cite{PengHerrmann:1994,PengHerrmann:1995,Poeschel:1994,Nagel:1995,HongEtAl:1994}.

In the present paper we are concerned with a two--dimensional cellular
automaton designed to simulate traffic in a city. As a special feature
of our model we emphasize the fact that we use rather realistic
traffic rules which will cause a slowing down of the average traffic
velocity and finally will lead to the collapse.

\section{The model}
\begin{figure}[ht]
\centerline{\psfig{figure=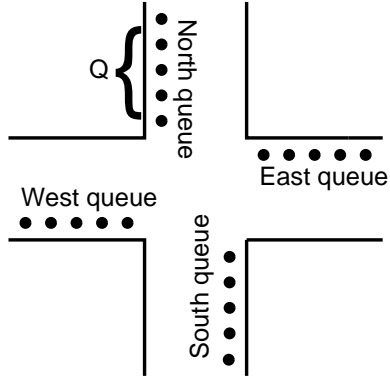,width=5cm,angle=270}}
\caption{The schematic plot of a crossing}
\label{fig:kreuzung}
\end{figure}

In our model the city is represented by a set of $L$ streets in
horizontal direction crossing $L$ streets in vertical direction. Cars
are allowed to move in both directions, i.e. we have no restrictions
to one--way streets. The crossings define a two--dimensional cellular
automaton, each of the $L^2$ crossings is represented by a site of the
automaton. We assume periodic boundary conditions in both directions.
Fig.~\ref{fig:kreuzung} shows a schematic plot of an automaton site (a
crossing). At each crossing there are four queues of maximum length
$Q$ filled by cars coming from one of the four directions,
representing the finite space where cars can move freely between
crossings in realistic urban traffic systems. Hence, at each crossing
there are allowed at most $4~Q$ cars.
Each of the cars has a desired direction, i.e. to the right, to the
left or straight on, which has to be chosen according to a certain
rule. We will discuss this point in detail below. Within each time
step the first of the cars in each queue can go to the next crossing
provided that it does not have to give way to one of the other cars at
the same site. In case it has to give way to another car it will not
move during this time step. In our model we assume the simple and
realistic rule that a car has to give way in the case that another car
occupies the same crossing at its right hand side. Moreover a car has
to stop if it intends to turn left and there is a car at the opposite
side of the crossing which goes straight on or turns to the right. If
all of the four top positions of a crossing are occupied by four cars
one of the cars will be selected randomly and will then be allowed to
move in the current time step. Insofar we have chosen realistic rules
which are current law in many countries.

Each of the $N$ cars in the system starts at a randomly selected
crossing, its desired direction (left, straight on, right) will be
determined according to some rule (discussed below). Within each time
step a car maximally moves a distance of one lattice cell, i.e. it
tries to transit to one of the next neighbors of the current site
provided the following three conditions hold:
\begin{enumerate}
\item It does not have to give way.
\item It is placed at the top position of its queue.
\item There are less than $Q$ cars standing in the destination queue.
\end{enumerate}
Otherwise it rests. Obviously, if there is only one car in town it
never has to give way nor will it find another car in front of itself.
Therefore the average velocity\footnote{Here and in the following we
describe the ensemble average of a value $x$ by the symbol $\langle x
\rangle$, whereas the time average by $\overline{x}$.}  $\langle v
\rangle =1$ in units of lattice cells per time steps.

As mentioned above we have to discuss the rule which determines the
desired direction within each time step. There are at least three
reasonable rules:
\begin{enumerate}
\item The desired direction for each car is selected at random within
  each time step not regarding whether the car moved in the previous
  time step or stopped. This rule is rather artificial since provided
  a jam occurs it can dissolve due to the fact that the cars might
  give up their previously desired direction when they cannot move.
\item Each car is assigned a destination site on the lattice. Once a
  car reached its destination it will be assigned a new randomly
  chosen destination. This rule is problematical since one has to
  define the detailed path a car follows to reach its destination. A
  natural choice would be to chose a path so that the car has to
  change its direction only once on its way between its starting
  position and its destination. In this case, however, the overall
  behavior of the system would depend on the size of the lattice $L$
  since one can check that the probability to be allowed to move
  straight on is higher than to be allowed to turn. Cars which intend
  to turn either left or right (on the average) have to give way more
  frequently than those which go straight on. The larger the system
  the smaller becomes the probability per time step to turn. For the
  case $L\rightarrow \infty$ the probability for a car in a given time
  step to turn approaches zero. Hence this rule does not allow for
  size--independent results, and the results of a simulation using
  this rule apply only for one fixed lattice size $L$.
\item Each car is assigned a desired direction during the
  initialization. But now, only after moving one step to the aimed
  destination a new desired direction will be chosen at random,
  i.e. the primarily chosen desired direction is maintained by each
  car until all of the three conditions hold which allow the car to
  move. This rule has several advantages compared with the other
  rules. First we avoid the size dependent behavior of the previous
  rule, and second we do not have long range correlations between the
  lattice sites for the case of low car density, which would be caused
  by the second rule. Hence the behavior of the system can be
  described by a Markov process, which will be the basic assumption of
  our analytical calculation in
  section~\ref{StatisticalDescription}. An analytical description
  which takes into account the long range correlation between the
  sites coming from the rules which determine the path of the cars
  seems to be very difficult, except perhaps for the (trivial) case
  $L\rightarrow \infty$, when the probability of the cars to turn, and
  hence the traffic rules connected with the turn, can be
  neglected. In the following we always refer to the third rule.
\end{enumerate}

\section{Numerical results}
\label{Numerics}
In this section we present the results found by
numerical simulations of a cellular automaton which represents the
traffic rules described above.

The two--dimensional automaton consists of $L\times L$ sites on a
rectangular lattice. To eliminate boundary induced effects we have
chosen periodic boundary conditions in both directions, i.e. we assume
the topology of a torus. We simulated systems of size $L=20$, $30$,
$50$, $100$ and no striking difference in the behavior of the system
was observed.  Hence, we exclude finite size effects. The fact that
already comparatively small lattices exhibit this independence from
their size seems to indicate the existence of only weak spatial
correlations between sites separated by a large distance. For high
densities, i.e.~close to or above the critical density $\eta_{cr}$,
this statement does not remain valid. In fact, we expect the existence
of long range correlations. However, in the present paper we are
mostly concerned with the low density region and we will not devote
ourselves extensively to the behavior of the automaton for high car
density.

Each of the $L\times L$ automaton sites had its own four queues of
length $Q=10$ connecting the site with one of its four neighboring
sites as described in the previous section. For each value of the
density $\eta=\frac{N}{L\times L}$, where $N$ is the number of cars,
we started the system at randomly chosen initial conditions. Each car
was assigned the initial site, the queue in this site (i.e. whether it
comes from North, East, South or West) and the desired direction
(left, straight, right) for the next step. When a car transits to an
empty site from a certain direction it occupies the first empty place
in the queue. If it is at the top position of the queue it can transit
to the next site or it has to wait, depending on the cars coming from
the other directions of the same site, and depending on the traffic
rules which apply to the given situation. If a car transits to another
site and if there are other cars located in the previously mentioned
queue, all of them advance by one position in this queue hence, moving
up ``closer to the crossing''.

After starting the simulation for a given number of cars, i.e. for a
given density $\eta$, we evolved the system 10,000 time steps without
recording the statistics in order to let the system forget about the
initial preparation. Typically a ``stationary'' behavior was found
after only a few hundred time steps. In density regions very close to
the transition value however, the relaxation of the system lasts
longer which hints at some phase--transition--like phenomenon. The
mean velocity of the cars was then computed using 2,000 time steps. We
varied the car density beginning from $\eta=0$ to $\eta=3.8$ in steps
$\Delta\eta=0.01$.

\begin{figure}[ht]
\centerline{\psfig{figure=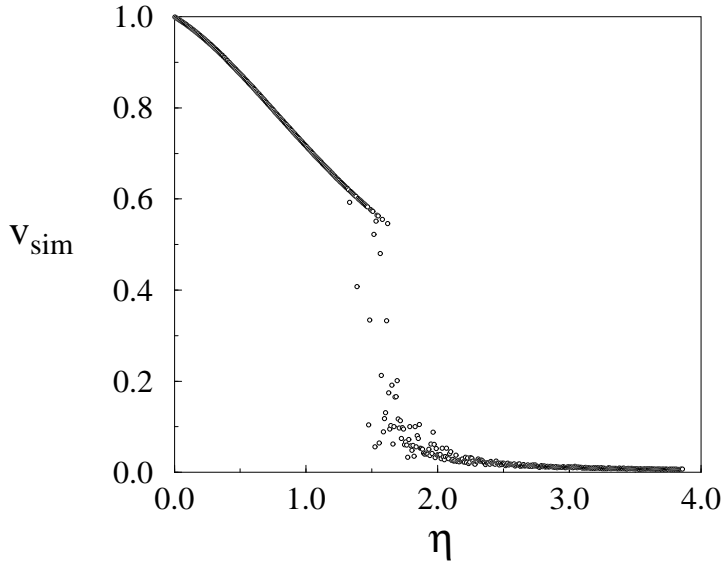,width=10.cm,angle=0}}
\caption{The simulation result: The mean velocity $v_{sim}$ of $N$
cars on a periodic lattice of size $L \times L$ vs the global traffic
density $\eta=N/\left(L\times L\right)$. Except for values around the
critical density $\eta_{cr}$ the curve is rather smooth. In the
vicinity of the transition zone critical fluctuations, i.e.~temporary
jams, cause an irregular relationship. This can be understood from the
fact that the typical lifetime of jams becomes of the same order as
the simulation time.}
\label{FIGsim}
\end{figure}

The results of the simulations are shown in fig.~\ref{FIGsim}. Like in
simulations of cellular automata performed by other authors with
different traffic rules we found a smoothly decaying function in the
low density region. At a certain threshold value $\eta_{cr}\approx
1.6$ the character of the traffic flow changes abruptly and the system
transits into the clogged regime. But different from other systems we
find a remnant velocity in the clogged regime. This is due to the rule
that in case all four of the queues are occupied there will be one car
selected randomly which then will be allowed to move. Therefore, in
our model there exist situations where a jamming at a crossing with
four queues occupied may still dissolve.

The curve in the small density region is very smooth, i.e. there are
almost no fluctuations. Hence, we can assume that long range
correlations do not play a major role there. In the following
section~\ref{StatisticalDescription} we will present a statistical
description appropriate to model the flow in the low density regime,
i.e.~for $\eta\ll \eta_{cr}$, which was inspired by this observation.

\section{A statistical description of vehicular traffic}
\label{StatisticalDescription}
Let the number of cars be denoted by $N$ and the number of cells
respectively crossings building the torus by $M=L\times L$ where $L$
is the common perimeter of the torus. For the global density we choose
the symbol $\eta=N/M$.

The quantity we focus our attention on is a time
averaged\footnote{When performing the simulation the parameter $T$ is
chosen sufficiently large in order to scan the state space
respectively the limit set with sufficient accuracy.} and normalized
velocity
\begin{equation}
  \overline{v} = {1\over T}\sum\limits_{t=1}^T{N-\Delta(t)\over N} = 1
  - {\langle\Delta \rangle_T\over N}~,
\label{TA}
\end{equation}
where $\Delta(t)$ means the total number of cars prevented from moving
by traffic rules at time $t$ which depends on the exact situation
experienced during the simulation.

 For a description of the dynamics in the low density region $\eta\ll 1$
we apply the following arguments:
\begin{enumerate}
\item An effective description is possible which separates the process
  of clustering at the $M$ possible crossings from the process of
  obeying the traffic rules. Due to this independence assumption the
  time average (\ref{TA}) extracted from a simulation can be
  decomposed into two averages: first a time and cluster average
  related to the traffic rules and second a time average with respect
  to the clustering process.
\item Given $i$ cars meeting at one crossing a variety of situations
  is possible. Every car is approaching the crossing from either
  North, East, South or West and intends to go straight, to turn left
  or to turn right. In the long run all possible situations will have
  equal probability. Hence, the average number of cars which must
  wait, denoted by $\overline{\delta_i}$ with
  $0\le\overline{\delta_i}\le i$, can be computed by combinatorial
  reasoning. The result of this computation is collected in
  table~\ref{waitingcars} for $i=1,\ldots,8$. A sketch of the
  derivation is given in appendix~\ref{AppendixA}.

For clusters of size $i>8$ the $\overline{\delta_i}$ are not shown
since their contributions are suppressed by very small cluster
probabilities.
\item Central to our description is the identification of cluster
  distributions, i.e.~we write $\underline{k}=(k_0,k_1,\ldots,k_N)$
  meaning that $k_i$ denotes the number of cells in the system where
  $i$ cars are meeting.  Of course, there are two boundary conditions,
  namely
\begin{equation}
  \sum\limits_{i=0}^N k_i=M
\label{BC1}
\end{equation}
and
\begin{equation}
  \sum\limits_{i=0}^N i\:k_i = \sum\limits_{i=1}^N i\:k_i = N
\label{BC2}
\; .
\end{equation}
We now assume that for the low density regime $\eta\ll 1$ the process
of clustering is ergodic which means for almost all initial
configurations the time average can equivalently be represented by an
average according to an invariant probability distribution. Note that
this probability distribution is defined over the space of all
configurations $\underline{k}$ compatible with the boundary conditions
(\ref{BC1}) and (\ref{BC2}). In order to derive this distribution we
additionally assume that the dynamics effectively is equivalent to a
process of independently distributing $N$ cars (balls) among $M$ cells
with crossings (urns) according to the equidistribution. This means we
assume the discrete dynamics to be a Bernoulli process. Then the
probability distribution reads
\begin{equation}
p(\underline{k}) 
\;=\; {M!~ N!\over M^{N}\prod\limits_{j=0}^N \left[(j!)^{k_j}~
  k_j!\right]} \; .
\label{understand}
\end{equation}
To understand formula~(\ref{understand}) in detail one has to realize
that there exist\\ $M!/\left[k_0! k_1!\ldots k_N!\right]$ different ways
to index the cells with numbers $0,1,\ldots, N$ since all cells
containing the same number $i$ of cars are not distinguishable. Then
there exist $N!/\left[ (0!)^{k_0} (1!)^{k_1}\ldots (N!)^{k_N}\right]$
ways to index the cars with numbers $1,\ldots,N$ taking care of the
fact that cars within an identical cell are not
distinguishable. Finally the equidistribution with respect to all
configurations yields the factor $M^{-N}$. This result was obtained by
von Mises (1939)~\cite{Mises:1939}.
\end{enumerate}
Now the time average $\langle\Delta\rangle _T$ is replaced by the
ensemble average
\begin{equation}
  \langle\Delta\rangle_{\underline{K}} \;=\;
  \sum\limits_{\underline{k}}^{\ast}
  p(\underline{k})\;\sum\limits_{i=1}^N 
    k_i\overline{\delta_i}\,\alpha_i \; .
\label{timeaverage}
\end{equation}
The sum with respect to $i$ reflects the average delay due to the
traffic rules at the crossings and the sum with respect to all
configurations obeying (\ref{BC1}) and (\ref{BC2}) -- which is
indicated by the symbol $\ast$ -- accounts for the cluster statistics.
The factor $\alpha_i$ accounts for the fact that the dynamics of the
real process differs from a Bernoulli--process.  In general there may
exist spatial as well as temporal correlations.  The spatial
correlations become important for densely filled lattices. The
dynamics which control the traffic within one cell is strongly
influenced by the cars moving in neighboring cells. Finally the
emergence of a traffic jam clearly expresses these correlations.  On
the contrary temporal correlations play the dominant role for very
sparse system, i.e.~in the low density region.  Consecutive situations
are not completely statistically independent. Let us briefly explain
this effect considering only cluster of size $i=2$: When the cars are
randomly distributed amongst the lattice sites there is a certain
probability that two cars $p$ and $q$ occupy the same site and hence
form a cluster of size $i=2$. Assuming statistical independence this
probability is time invariant but for the cellular automaton the
situation is quite different.

If two cars meet at a site there exist two possibilities: either both
cars continue, or one of them continues and the other one has to give
way. There is no chance that both cars meet again one time step
later. In the extreme case of only two cars, i.e. $N=2$, the
probability for them to meet by random distribution is thus
effectively reduced by a factor of two in comparison with the
automaton dynamics. Given only a small number $N$ of cars, i.e
$\eta=N/M\ll 1$, and provided only two of them $p$ and $q$ meet
occasionally one has to consider in the following time step a system
where only $N-2$ cars feel the interaction with (in our case: can meet
occasionally) $N-1$ other cars and $2$ cars which can meet only $N-2$
cars, i.e. $p$ is not allowed to meet $q$ and $q$ cannot meet
$p$. Hence, the system has some kind of memory and therefore it is not
ideally Bernoullian.

\begin{figure}[ht]
\centerline{\psfig{figure=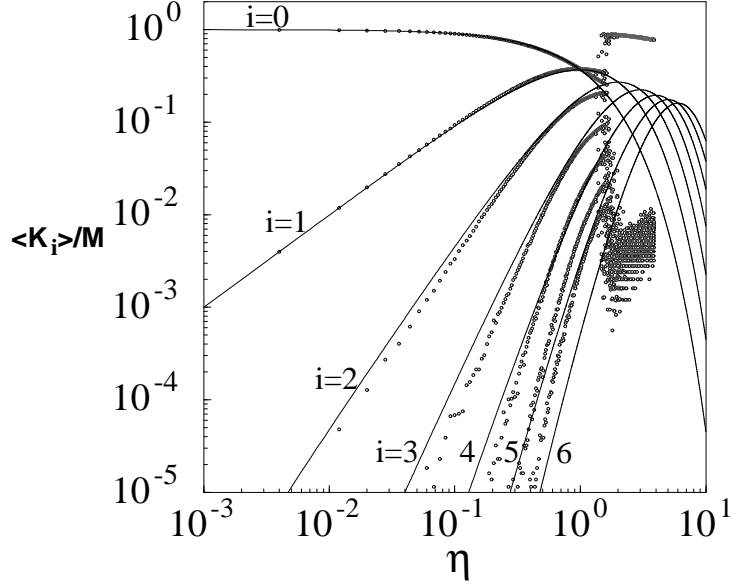,width=10.cm,angle=0}}
\caption{The average number of clusters of size $i$ normalized to the
lattice size $M=L \times L$ as a function of the global traffic
density $\eta=N/M$. The lines represent
eq.~(9) while the points correspond to
numerical data achieved by a simulation on a $50\times50$ lattice
(torus). The curve $i=0$ gives the probability for a site to be
empty.}
\label{FIGcluster}
\end{figure}

Obviously this effect plays a crucial role only for small $N$ since
otherwise the relative difference between $N-1$ or $N-2$ is
negligible. Speaking in terms of density this means that the effect is
substantial only for very small densities. Then however, clusters
larger than $i=3$ are suppressed by small likelihood. The net effect
of these correlations can be taken into account through counting all
possible configurations which can be realized after two cars have
met. This procedure is similar to the method employed for the
determination of the $\overline{\delta}_i$ (as explained in
appendix~\ref{AppendixA}). Neglecting contributions related to
clusters of size larger than three, which is safe for very small
densities (see fig~\ref{FIGcluster}, explanation below) means
$\alpha_i=1$ for $i\ge 4$. From simple geometric considerations we
find $\alpha_1=1$, $\alpha_2=1/2$ and $\alpha_3=7/4$.

Inserting expression~eq.(\ref{timeaverage}) into (\ref{TA}) yields
\begin{equation}
  \overline{v} = 1 - {\langle\Delta\rangle_{\underline{K}} \over N} =
  1 - \sum\limits_{i=1}^N\: {\overline{\delta_i}\alpha_i\over N}
  \sum\limits_{k_1,\ldots,k_N} ^{\ast}\: k_i \: p(k_1,\ldots,k_N)
\end{equation}
and performing the sum over all $k_j$ ($j\ne i$) we find
\begin{eqnarray}
  \overline{v} &=& 1 - \sum\limits_{i=1}^N\:
    {\overline{\delta_i}\alpha_i\over N}\:\sum\limits_{k_i=0}^{\lfloor
    N/i\rfloor} k_i\: p(k_i)\nonumber\\ &=& 1 - \sum\limits_{i=1}^N\:
    {\overline{\delta_i}\alpha_i\over N}\; \langle K_i \rangle
\label{moment}
\; .
\end{eqnarray}
The distribution $p(k_i)$ -- declared for $i=0,1,\ldots,\lfloor
N/i\rfloor$ -- can be calculated using the inclusion--exclusion
principle~\cite{JohnsonKotz:1977,Krengel:1990}. This derivation is
performed in appendix~\ref{AppendixB} and the result reads
\begin{equation}
  p(k_i) = {N!\over M^N}\:\sum\limits_{j=k_i}^M\:
  (-1)^{(j-k_i)}\left({j\atop k_i}\right)\: {(M-j)^{(N-ji)}\over
  (i!)^l~ (N-ji)!}~.
\label{pvonKi}
\end{equation}
We see from (\ref{moment}) that the ingredients we actually need are
the first moments $\langle K_i \rangle$ of the cluster distribution.
They can be derived calculating a generating function for the
(descending factorial) moments, denoted
$H^{(M)}_i(z,x)$~\cite{JohnsonKotz:1977}. This calculation is
performed in appendix~\ref{AppendixC}. The result reads
\begin{equation}
  \langle K_i \rangle = \left({N\atop i}\right)\; {1\over M^{(i-1)}}\;
  \left(1-{1\over M}\right)^{(N-i)}~.
\label{momentsofclusterdistr}
\end{equation}

In fig.~\ref{FIGcluster} we have depicted the expectation value of the
cluster distribution $\langle K_i \rangle$
(eq.~(\ref{momentsofclusterdistr})) divided by the system size
$M=L\times L$. The solid lines correspond to the function given in
eq.~(\ref{momentsofclusterdistr}) and the points represent related
data taken from our simulation on a $50\times 50$ torus.  The values
for $i=0$ give the probability for a site to be empty. For $i=0$ and
$i=1$ the analytical and numerical results agree perfectly, for larger
$i$ the discrepancies between the solid lines and points are a direct
consequence of the deviations from the assumed independent behavior as
explained above. Note that the probabilities in fig.~\ref{FIGcluster}
are plotted using a logarithmic scale. The observed discrepancies for
clusters of size larger than 3 will hardly have any influence on the
overall behavior of the automaton. For values $\eta>\eta_{cr}$ the
simulation data illustrate a breakdown for the small occupation
numbers (except $i=0$) due to the emergence of jams which are clusters
of high order. At the transition point the occupation rates abruptly
drop by three orders of magnitude. The precise position of this
transition point is masked by the finite size of $N$ used in the
simulation. Increasing the car density $\eta$ beyond the blurred
critical zone results in a slow rise of the occupation number for the
small clusters and simultaneously, in a slow decline of the number of
empty sites. This is due to the increasing number of cars still moving
around the jammed crossings.

\begin{figure}[ht]
\centerline{\psfig{figure=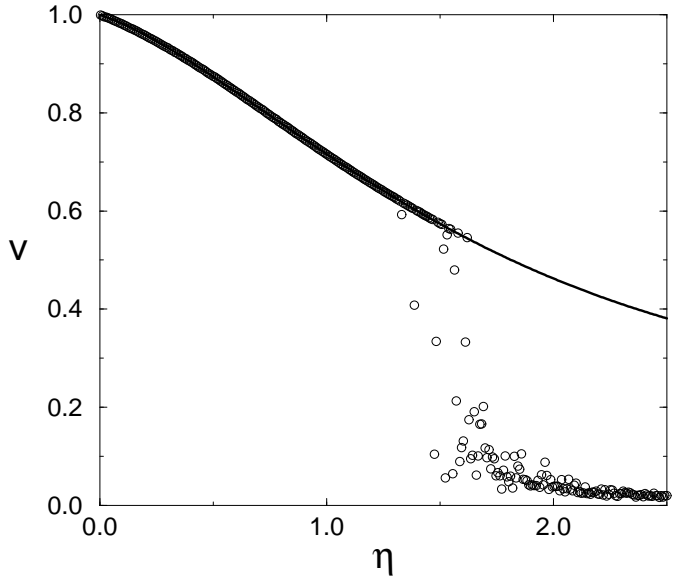,width=10.cm,angle=0}}
\caption{The average velocity vs the global traffic density for the
low density regime (non jamming traffic). The curve displays the
result from 
equation~(11),
the points show data taken from the
simulation.}
\label{FIGEQFF}
\end{figure}

Inserting equation~(\ref{momentsofclusterdistr}) into (\ref{moment})
yields the following formula
\begin{equation}
  \overline{v} = 1 - \sum\limits_{i=1}^N\:
    {\overline{\delta_i}\alpha_i\over N}\; \left({N\atop i}\right)\;
    {1\over M^{(i-1)}}\; \left(1-{1\over M}\right)^{(N-i)}~.
\label{EQF}
\end{equation}
The last step will be to substitute in (\ref{EQF}) the density $\eta$
for the number of cars according to $N=\eta M$ which results in
\begin{eqnarray}
  \overline{v} = 1 &-& \sum\limits_{i=1}^{\eta M}\:
  {\overline{\delta_i}\alpha_i\over i!}\; \left(\eta-{1\over M}\right)
  \cdot\ldots\nonumber\\
&& \cdot\left(\eta-{i-1\over M}\right) ~\left(1-{1\over
    M}\right)^{(\eta M-i)}
\label{EQFF}
\end{eqnarray}
The main result of our statistical description, namely the mean
velocity as a function of the car density (eq.~(\ref{EQFF})), is
plotted in fig.~\ref{FIGEQFF}.  The solid line shows the function
given by eq.~(\ref{EQFF}) and the points are the data taken from the
simulation on the $50\times 50$ torus (compare fig.\ref{FIGsim}. In
the low density region we find a nice agreement. The closer one
approaches the transition point the more eq.~(\ref{EQFF})
overestimates the simulation results. This can be understood by
recalling the fact that for densities close to the critical value the
effective decomposition of the automaton dynamics into the cluster
dynamics and the single site dynamics becomes inappropriate due to
formation and dissolution of short lived jams (like critical
fluctuations).  They tend to increase the occupation number of larger
clusters and hence, lead to an effective decrease of the average
velocity.

\begin{figure}[ht]
\centerline{\psfig{figure=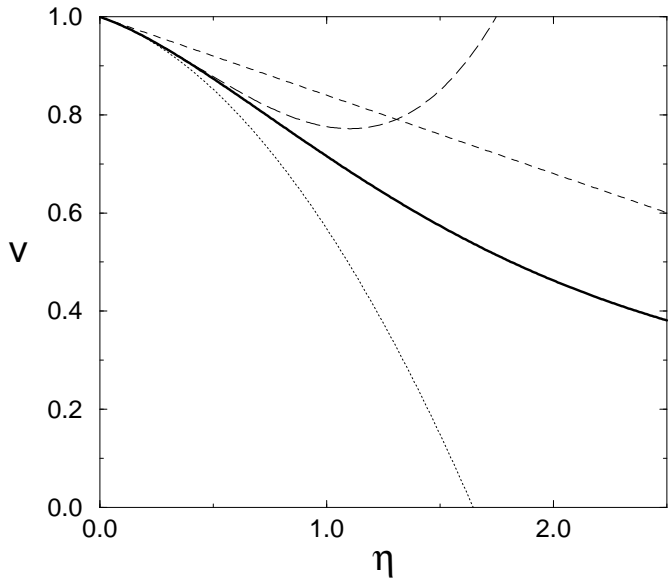,width=10.cm,angle=0}}
\caption{The average velocity as derived from the statistical
description~(eq.~(9)) 
(fat line) and
truncated series expansions: linear (dashed), quadratic (dotted) and
cubic (long dashed). This plot clearly illustrates a nonlinear
relationship between average velocity and global density.}
\label{FIGexpansion}
\end{figure}

In the thermodynamic limit -- which means $M,N\to\infty$ and keeping
$\eta=N/M$ constant -- we drop terms of order ${\cal O}(M^{-1})$ and
end up with
\begin{equation}
  \overline{v} \stackrel{M\to\infty}{\approx} 1 -
  \sum\limits_{i=1}^{\eta M}\: {\overline{\delta_i}\alpha_i\over i!}\;
  \eta^{(i-1)}\;\exp(-\eta)
\end{equation}
where the exponential corresponds to the dominant contribution of the
last factor in (\ref{EQFF}). This formula can be expressed by a series
expansion in powers of $\eta$
\begin{eqnarray}
  \overline{v} \stackrel{M\to\infty}{\approx} &1& -
  \;\sum\limits_{i=1}^{\eta
  M}\:\sum\limits_{k=0}^{\infty}\:{(-1)^k\overline{\delta_i}\alpha_i\over
  i!  k!}\:\eta^{(i-1+k)}\label{expansion}\\ =\;&1& \!\!-\!\!
  \;\left({\overline{\delta_2}\alpha_2\over 2}\right)\eta \:\!\!+\!\!\:
  \left({\overline{\delta_2}\alpha_2\over 2} -
  {\overline{\delta_3}\alpha_3\over 6}\right) \eta^2-
  \left({\overline{\delta_2}\alpha_2\over
  4}-{\overline{\delta_3}\alpha_3\over
  6}+{\overline{\delta_4}\alpha_4\over
  24}\right)\eta^3\pm\ldots\nonumber
\end{eqnarray}
Note that there is no constant term in the double sum since
$\overline{\delta_1}=0$.  Equation (\ref{expansion}) can be truncated
at a given order of $\eta$ hence, giving rise to an approximation
scheme. In fig.~\ref{FIGexpansion} we plotted the linear (dashed), the
quadratic (dotted) and cubic (long dashed) approximations together
with the full formula (fat solid). The range of reliability visibly
increases with increasing the order of approximation. On the other
hand this scheme clearly illustrates that the functional relation
between mean velocity and global density definitely is {\em
nonlinear}, except for very small car densities.

\section{Conclusion}
We investigated the behavior of a two--dimensional cellular automaton
with periodic boundary conditions to simulate traffic flow in
cities. The automaton mimics realistic traffic rules which apply to
our everyday experience of vehicular traffic. In agreement with
similar models we found a slow decay for the mean velocity $\langle v
\rangle$ as a function of the global traffic density $\eta$.  At a
critical threshold value $\eta_{cr}$ the mean velocity collapses
abruptly and the system transits into another regime of global
behavior which we call the jamming regime. Beyond the critical density
$\eta_{cr}$ the average velocity is very small and it declines further
when increasing the density. By simulating different sizes of automata
we could exclude the influence of finite size effect provided the
density value is not rather close to the critical density.

Applying combinatorics and statistical methods for the description of
the system in the low density regime the analytical calculation
performed in section~\ref{StatisticalDescription} yielded the average
car velocity as a function of the car density --
$\langle\overline{v}\rangle\left(\eta\right)$ -- which nicely agrees
with the values of the numerical simulations in
section~\ref{Numerics}.  Since the derived functional relationship
between mean velocity and global traffic density was based on very
general assumptions (conservation of the number of cars, weak spatial
correlations in the low density regime, ergodicity) we expect this
description to be valid for a wider class of traffic systems.  Note
that the specific structure of traffic rules enter the description
only through the average delays $\delta_i$. The only nonlocal
ingredient arises from the restriction that cars have to stop in case
the next desired queue at a neighboring site is totally filled up. But
this nonlocal character only becomes substantial for densities close
to the critical value where it causes long range spatial correlations.

Several authors (e.g.~\cite{MartinezCuestaMoleraBrito:1995}) assert
that the average velocity $\overline{v}$ in the low density regime is
a linear function of the density $\eta$ and indeed the simulation
results seem to support this observation. A more detailed analysis
however reveals that this function definitely is nonlinear. An
analytic expansion shows (eq.~(\ref{expansion}),
fig.~\ref{FIGexpansion}) that the results become dramatically wrong
when truncating the formula after the linear term.  Surprisingly the
analytic description gives good results even for densities not too
small where the nonlocal effects of the dynamics and hence, long range
correlations spoil the basic assumptions of our treatment.

For still higher values of the density $\eta$ the results of our
simulation agree with results reported in literature
(e.g.~\cite{MartinezCuestaMoleraBrito:1995} and many others), 
but no longer with our approximate theory.  Above a
critical density we observe an abrupt transition of the system into
the jammed regime where the averaged velocity is close to zero.
This regime however is beyond the scope of our statistical description
and has to be investigated starting from other approaches
(e.g.~nucleation processes).

\ack
We thank H.~Herzel and L.~Schimansky--Geier for helpful discussion. We
greatly acknowledge U.~K\"uchler for drawing our attention to
ref.~\cite{JohnsonKotz:1977} and for discussing the combinatorics in
detail.

\appendix
\section{Calculation of the average number of waiting
cars $\overline{\delta_i}$ for clusters of size $i=1$ and $i=2$}
\label{AppendixA}

In table~\ref{waitingcars} we present the average number $\delta_i$ of
cars which are prevented from moving given $i$ cars meeting at a
crossing. Obviously the number $\delta_i$ is confined to the interval
$[0,i]$. 

\begin{table}[ht]
\caption{The average number of cars which are stopped
$\overline{\delta_i}$ when $i$ cars meet at a crossing.}
\label{waitingcars}
\begin{center}
\begin{tabular}{c c} 
\#cars meeting $i$ &$\overline{\delta_i}$\\
 \hline
1&0.0\\
2&0.638889\\
3&1.479167\\
4&2.467014\\
5&3.524740\\
6&4.604709\\
7&5.683838\\
8&6.752964\\ 
\end{tabular}
\end{center}
\end{table}

The average is performed with respect to all possible
configurations which are assumed to possess equal probability.  The
case $i=1$ is rather simple: If there is only one car at the crossing
it has never to give way hence, $\overline{\delta_1}=0$.  The case
$i=2$ is not that trivial and the computation of $\overline{\delta_2}$
requires some combinatorial effort. Since each of the cars can wait at
one of the four sides of the crossing and can proceed in one of the
three directions (left, straight on, right) we have $4\times 3 \times
4 \times 3=144$ different situations of equal probability. Each of
these situation can be assigned to one of the situations in the left
column of table~\ref{situations}. We now explain the columns of
table~\ref{situations}:
\begin{table}[ht]
\caption{All situations which might occur when two cars meet at a
crossing, their frequency, their symmetry according to rotation and due
to indistinguishability of the cars and the number of cars which have
to stop in the current situation (explanation in the text).}
\begin{center}
\begin{tabular}{c c c c c c}
  graph & rot.~symm. & dist. & \#events & waiting &\#stopping\\ 
  \hline
  \psfig{figure=vorfahrt1.eps,width=0.5in,angle=0} & 
     $4\times 3 \times 3$ & 1 & 36 & 1 & 36\\
  \psfig{figure=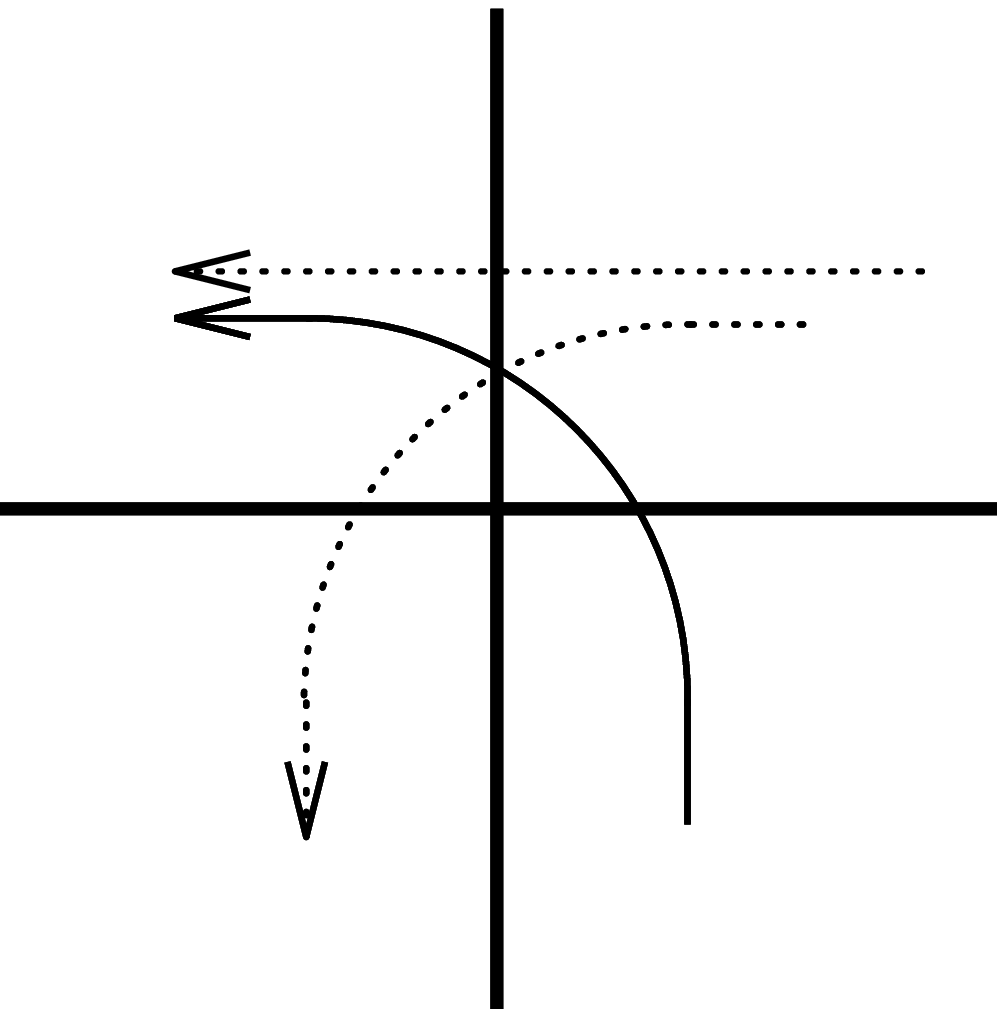,width=0.5in,angle=0} & 
     $4\times 1 \times 2$ & 2 & 16 & 1 & 16\\ 
  \psfig{figure=vorfahrt3.eps,width=0.5in,angle=0} & 
     $4\times 1 \times 1$ & 2 & 8  & 0 &  0\\
  \psfig{figure=vorfahrt4.eps,width=0.5in,angle=0} & 
     $4\times 1 \times 3$ & 2 & 24 & 1 & 24\\ 
  \psfig{figure=vorfahrt5.eps,width=0.5in,angle=0} & 
     $4\times 1 \times 3$ & 2 & 24 & 0 &  0\\
  \psfig{figure=vorfahrt6.eps,width=0.5in,angle=0} & 
     $2\times 1 \times 1$ & 2 &  4 & 0 &  0\\
  \psfig{figure=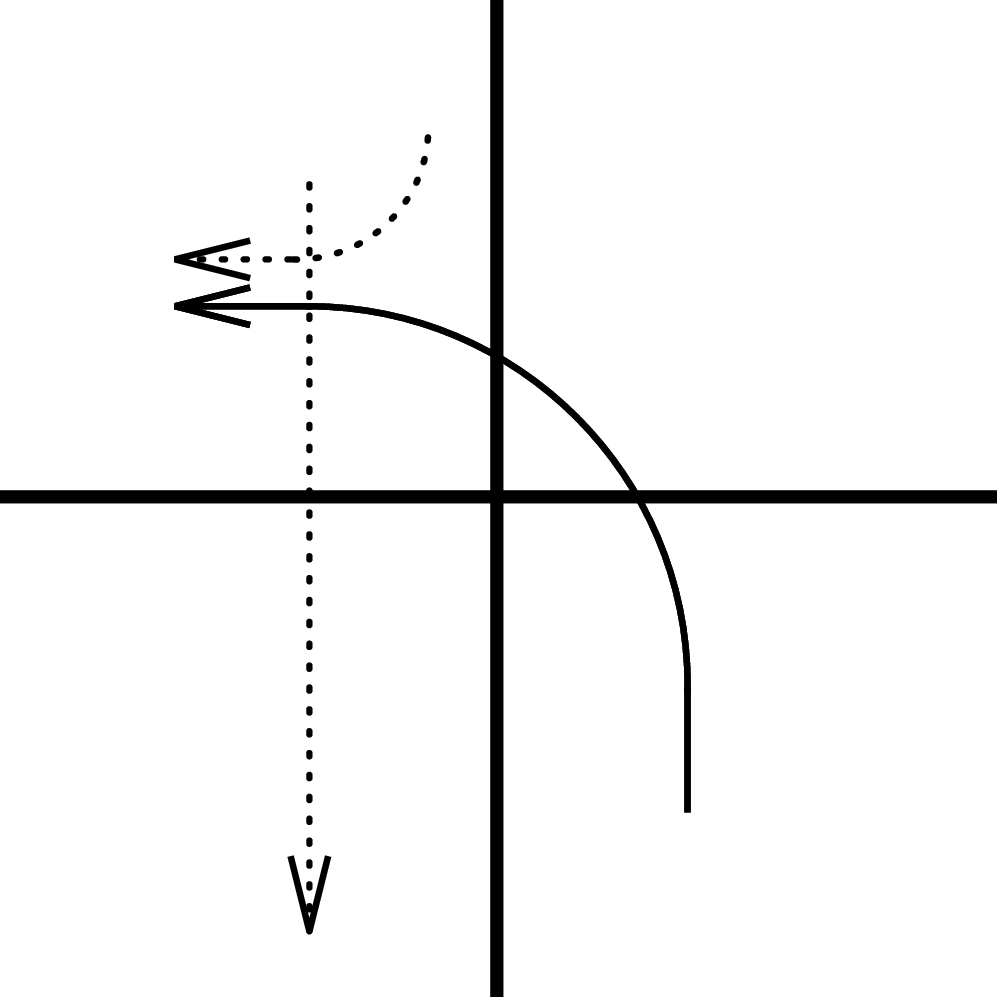,width=0.5in,angle=270} & 
     $4\times 1 \times 2$ & 2 & 16 & 1 & 16\\
  \psfig{figure=vorfahrt8.eps,width=0.5in,angle=0} & 
     $2\times 1 \times 1$ & 2 &  4 & 0 &  0\\
  \psfig{figure=vorfahrt9.eps,width=0.5in,angle=0} & 
     $4\times 1 \times 1$ & 2 &  8 & 0 &  0\\
  \psfig{figure=vorfahrt10.eps,width=0.5in,angle=0} & 
     $2\times 1 \times 1$ & 2 &  4 & 0 &  0\\
  \hline
  & total&& 144 && 92\\
\end{tabular}
\end{center}
\label{situations}
\end{table}

\begin{enumerate}
\item A sketch of the configuration given by a related graph.
\item The number of the different realizations that relate to the
graph in the first column assuming first that the cars are
distinguishable. The first factor $A$ in the form $A\times B\times C$
originates from the symmetry according to rotation by
$\frac{1}{2}\pi$, $\pi$ and $\frac{3}{2}\pi$. The second factor $B$
denotes the number of choices for the first car (which is now assumed
to come from South) and the last factor $C$ gives the number of
situations possible for the second car.
\item When we take into account that the cars in fact cannot be
distinguished the number of different events that belong to the figure
in the first column has to be multiplied by the factor given in the
third column.
\item The total number of events for the whole class denoted by the
graph, i.e.~(column $2 \times$ column $3$).
\item The number of cars stopped by the traffic rules at this crossing
in one of the possible realizations (either one or none).
\item The total number of cars which have to stop related to the whole
class denoted by the graph, i.e.~(column $4 \times$ column $5$).
\end{enumerate}

From the last line in table~\ref{situations} one sees that there are
144 different situations of equal probability which amounts to 288
cars.  Furthermore we see that 92 of those 288 cars have to stop.
Hence, the average number of the cars stopped for 2--clusters is
$\overline{\delta_2}=92/144\approx0.639$.

For the clusters of size $i=3$ the average number of cars which are
prevented from moving $\overline{\delta_3}$ can be calculated in
analogy however the corresponding table contains about 8 times as much
situations as table~\ref{situations}. Therefore we do not want to
present it here. The result is $\overline{\delta_3}=1.479$. For cluster
of size larger than 3 the corresponding $\overline{\delta_i}$ have
been calculated by the method of complete enumeration using a
computer. These empirical results are collected in
table~\ref{waitingcars}. For $i=1\dots 3$ they are identical with the
analytical results.

For completeness sake we remark that for large cluster sizes ($i>8$)
there exists a negligible probability that one of the directions of
the crossing is not occupied. Typically there are many cars waiting in
each of the four queues. In those very likely cases our traffic rules
allow exactly one (randomly chosen) cars to drive, all other $i-1$
cars have to stop. Therefore we find
$\overline{\delta_i}\stackrel{i\gg 1}{\rightarrow} i-1$. Certainly the
contributions of such big clusters will only play a minor role in our
calculations due to very small likelihood.

\section{The derivation of the cluster probabilities}
\label{AppendixB}

The probability $p\left(k_i\right)$ to find exactly $k$ sites which
are occupied each by $i$ (independently moving) cars in a system of
$M$ crossings and $N$ cars is given in eq.~(\ref{pvonKi}). The
derivation of this formula employing the inclusion--exclusion
principle will be performed in this appendix. Note that the problem to
find $p\left(k_i\right)$ is different from the trivial problem to find
the probability for the $k_i$ crossings which are occupied by {\em at
least} $i$ cars.

The inclusion--exclusion principle relates the probabilities for a
finite number of sections of events to the probabilities of an exact
number of events.  \vspace{1.cm}

Let $(\Omega,P)$ be a probability space, $A_1,\ldots,A_N$ be events,
and for arbitrary $\{m_1,\ldots,m_j\}\subset \{1,\ldots,N\}$ let
$P(A_{m_1}\cap\ldots\cap A_{m_j})$ be known probabilities.  We define
\begin{equation}
  S_j= \sum\limits_{m_1,\ldots,m_j}\; P(A_{m_1}\cap\ldots\cap
  A_{m_j})
\end{equation}
Moreover, let be $B_n=\{\omega\in\Omega:\omega\in A_m$ for exactly $n$
values of $k\}$. Then the inclusion--exclusion principle asserts that
\begin{equation}
  P(B_n)=\sum\limits_{j=n}^N\; (-1)^{(j-n)}\:\left({j\atop n}\right)
  S_j
\label{INCLEXCL}
\end{equation}

Applied to our case the $S_j$ are the probabilities that $j$ cells
each contain $i$ cars and the rest of the cars, i.e.~$(N-ji)$ cars,
are distributed arbitrarily among the remaining $(M-j)$ cells; this
probability can be derived with ease and reads
\begin{equation}
  S_j= {N!\over (i!)^j \: (N-ji)!}\;\left( {1\over
    M}\right)^{(ji)}\;\left( 1-{j\over M}\right)^{(N-ji)}
\end{equation}
To explain this formula we first index each of the $N$ cars with the
numbers $1,\ldots,N$ which yields the factor $N!$. Since the $i$ cars
within each of the $j$ cells are not distinguishable we have to divide
this factor by $(i!)^j$ and since the remaining $(N-ji)$ cars are not
distinguishable too additionally by $(N-ji)!$. After having numbered
the cars we subsequently fill the first cell with cars $1,\ldots,i$,
the second cell with cars $(i+1),\ldots,(2i)$ and so on. In this way
we distribute the cars numbers $1,\ldots,ji$ among the urns
$1,\ldots,j$ yielding the factor $M^{(-ji)}$. The remaining $(N-ji)$
cars are distributed successively among the remaining $M-j$ cells at
random which explains the factor $\left[(M-j)/M\right]^{(N-ji)}$.

Insertion of this probability into the inclusion--exclusion formula
(\ref{INCLEXCL}) yields
\begin{eqnarray}
  p(k_i) &=& \sum\limits_{j=k_i}^M\: (-1)^{(j-k_i)}\left({j\atop
    k_i}\right) S_j \nonumber\\[.2cm] &=& {N!\over
    M^N}\:\sum\limits_{j=k_i}^M\: (-1)^{(j-k_i)}\left({j\atop
    k_i}\right)\: {(M-j)^{(N-ji)}\over (i!)^j~ (N-ji)!}
\end{eqnarray}
Because of the generalized definition of factorials -- using the gamma
function -- the sum effectively only ranges from $k_i$ up to $\lfloor
M/i\rfloor$ which is sensible.

\section{Calculation of the first moments of the cluster
distribution $\langle K_i \rangle$}
\label{AppendixC}

In the following we will derive the mean value of the cluster
distribution $\langle K_i \rangle$ which is given in
eq.~(\ref{momentsofclusterdistr}).

The generating function for the (descending) factorial moments of the
cluster distribution we use is
\begin{equation}
  H^{(M)}_i(z,x)=\sum\limits_{N=0}^{\infty}\;
  \sum\limits_{k_i=0}^{\infty}\:{M^N z^N\over N!}\:x^{k_i}\:p(k_i,N)
\end{equation}
where $p(k_i,N)$ is the probability to find a value $k_i$ for the
stochastic variable $K_i$ when trying with $N$ balls. The benefit of
such a complicated looking generating function is that the sums can be
performed yielding an analytical expression namely
\begin{equation}
  H^{(M)}_i(z,x)= \left[ \exp^z +{z^i\over i!}(x-1)\right]^M
\end{equation}
For the explicit derivation the reader is referred to the book by
Johnson and Kotz\footnote{one has to identify $j\equiv i, m\equiv M,
n,\equiv N, M_i\equiv K_i, g\equiv k_i, \mbox{Pr}[M_j=g|n]\equiv
p(k_i)$} p.~116ff~\cite{JohnsonKotz:1977} .  Moreover, this function
is related to the (descending) factorial moments of the stochastic
variable $K_i$ according to
\begin{eqnarray}
  {N!\over M^N}&&\; \left.\left({d^r\over dx^r}\;
H^{(M)}_i(x,z)\right)\right|_{x=1}
\nonumber\\
  &&= \sum\limits_{N=0}^{\infty} z^N\;\sum\limits_{k_i=0}^{\infty}
  k_i\:(k_i-1)\:\ldots(\:k_i-r+1)\;p(k_i,N)\nonumber\\ &&=
  \sum\limits_{N=0}^{\infty} z^N\;\; \Big\langle {k_i
    (k_i-1)\ldots\:(k_i-r+1)}\Big\rangle
\end{eqnarray}
hence, the first moment $\langle K_i \rangle$ is given as the
coefficient of $z^N$ in
\begin{equation}
  {N!\over M^N}\;{d\over dx}\left( H^{(M)}_i(z,x)\right|_{x=1}
\end{equation}
Consequently we calculate
\begin{eqnarray}
  {N!\over M^N}\left.\left({d\over dx}\left[ \exp^z +{z^i\over
    i!}(x-1)\right]^M \right) \right|_{x=1}
&=& {N!\over M^{(N-1)} }\:
  \exp^{\left[z(M-1)\right]}\: {z^i\over i!}\nonumber\\[.2cm] &&\!\!\!\!\!\!\!\!\!\!\!\!\!\!\!\!\!=
  \sum\limits_{k=0}^{\infty}\; {N!\over i!\; k!}\; {(M-1)^k\over
    M^{(N-1)}}\; z^{(k+i)}
\end{eqnarray}
and finally arrive at
\begin{equation}
  \langle K_i \rangle = {N!\over i!\; (N-i)!}\; {(M-1)^{(N-i)}\over
    M^{(N-1)}}= \left({N\atop i}\right)\; {1\over
    M^{(i-1)}}\; \left(1-{1\over M}\right)^{(N-i)}
\end{equation}

\end{document}